\documentclass{SciPost}

\usepackage{amsmath}
\usepackage{ulem}
\usepackage{bm}
\usepackage{amssymb}
\usepackage{physics}
\usepackage{tikz}
\usetikzlibrary{patterns}
\usetikzlibrary{arrows.meta}
\usetikzlibrary{decorations.pathmorphing}
\usepackage{orcidlink}
\usepackage{float}
\hypersetup{
    colorlinks,
    linkcolor={red!50!black},
    citecolor={blue!50!black},
    urlcolor={blue!80!black}
}

\usepackage[bitstream-charter]{mathdesign}
\DeclareSymbolFont{usualmathcal}{OMS}{cmsy}{m}{n}
\DeclareSymbolFontAlphabet{\mathcal}{usualmathcal}

\fancypagestyle{SPstyle}{
\fancyhf{}
\lhead{\colorbox{scipostblue}{\bf \color{white} ~SciPost Physics }}
\rhead{{\bf \color{scipostdeepblue} ~Submission }}

\fancyfoot[C]{\textbf{\thepage}}
}

\begin{document}

\pagestyle{SPstyle}

\begin{center}{\Large \textbf{\color{scipostdeepblue}{
Comment on "Time crystals made of electron-positron pairs"
}}}\end{center}

\begin{center}\textbf{
Zhen-Hua Feng\textsuperscript{1$\star$}\orcidlink{0009-0000-2744-2368},
}\end{center}

\begin{center}
{\bf 1} School of Physics and Astronomy, Beijing Normal University, Beijing 100875, China
\\[\baselineskip]
$\star$ \href{mailto:email1}{\small fengzhenhua@mail.bnu.edu.cn}
\end{center}

\section*{\color{scipostdeepblue}{Abstract}}
\textbf{\boldmath{%
Direct reproduction of Bialynicki-Birula's quantum solutions using the authors' own equations and initial conditions reveals two fundamental flaws. First, the eigenfunctions exhibit divergence in the region $y<0$, contradicting the claimed decay behavior ($\abs{\phi}\to\infty$ as $y\to -10$) (Figs. 9--11). Second, the apparent parity symmetry displayed in Figures 9--11 is not supported by Equation (21); numerical solutions clearly show asymmetric wavefunctions. Furthermore, all quantum solutions presented in Section IV violate the boundedness requirements of quantum mechanics. These inconsistencies raise serious concerns about the physical validity of the quantum framework underlying the time crystal model.
}}

{\bf Data availability:} All Python codes for numerical reproduction are available at \url{https://github.com/fungzhenhua/PythonSource}.

\vspace{\baselineskip}

\noindent\textcolor{white!90!black}{%
\fbox{\parbox{0.975\linewidth}{%
\textcolor{white!40!black}{\begin{tabular}{lr}%
  \begin{minipage}{0.6\textwidth}%
    {\small Copyright attribution to authors. \newline
    This work is a submission to SciPost Physics. \newline
    License information to appear upon publication. \newline
    Publication information to appear upon publication.}
  \end{minipage} & \begin{minipage}{0.4\textwidth}
    {\small Received Date \newline Accepted Date \newline Published Date}%
  \end{minipage}
\end{tabular}}
}}
}




\section{Contradiction in Eigenfunction Behavior Between Reproduction and Original Figs. 9--11 in Ref. [1]}

In the work of Ref. \cite{bialynicki-birulaTimeCrystalsMade2021}, the authors state in Sec. IV ("Quantum Hamiltonian," p. 5) that "All eigenfunctions seem to exhibit a similar slow falloff with increasing |y| and the spacing between maxima decreases with the increase of the energy." However, upon reproducing Figs. $9-11$ using the exact parameters specified in the original manuscript , we observed divergent behavior in the $y<0$ region for the eigenfunctions, which contradicts the results presented in the corresponding figures of Ref. \cite{bialynicki-birulaTimeCrystalsMade2021}.

In the concluding remarks on p. 5 of Ref. \cite{bialynicki-birulaTimeCrystalsMade2021}, the authors state: "At the quantum level we found the eigenstates of the quantum Hamiltonian operator." However, our numerical results do not support this significant claim.

\textbf{Why this claim matters:} 
The existence of bounded eigenstates is a fundamental requirement for any quantum-mechanical description of time crystals. If the reported eigenfunctions diverge or artificially impose unphysical symmetries, then the theoretical foundation for the entire model regarding quantum time crystals will collapse completely. We therefore meticulously reproduce the eigenfunctions using the same equations and initial conditions as Ref. \cite{bialynicki-birulaTimeCrystalsMade2021}, exposing two critical inconsistencies below.

\begin{figure}[H]
    \centering
    \includegraphics[height=3cm]{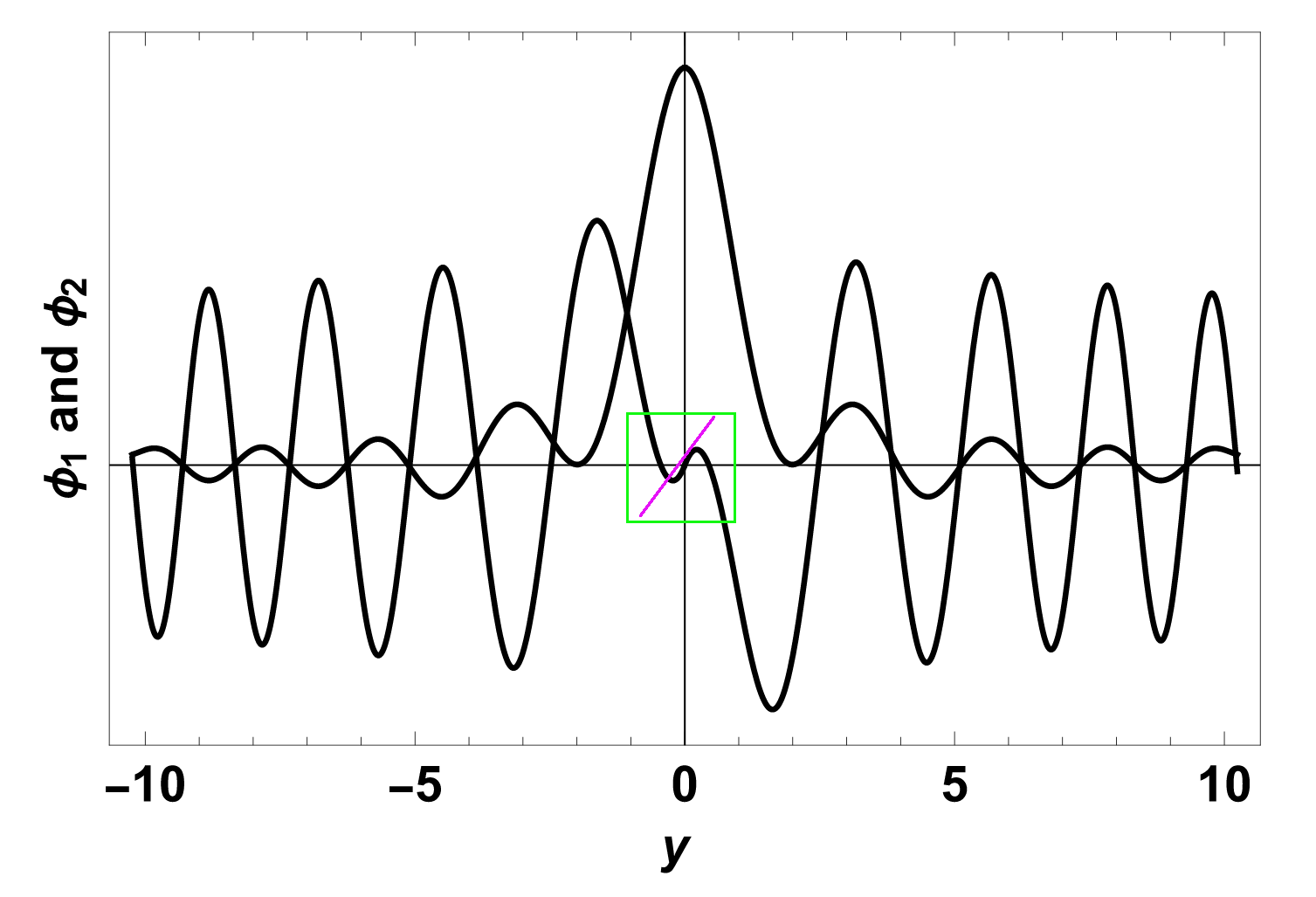}
    \includegraphics[height=3cm]{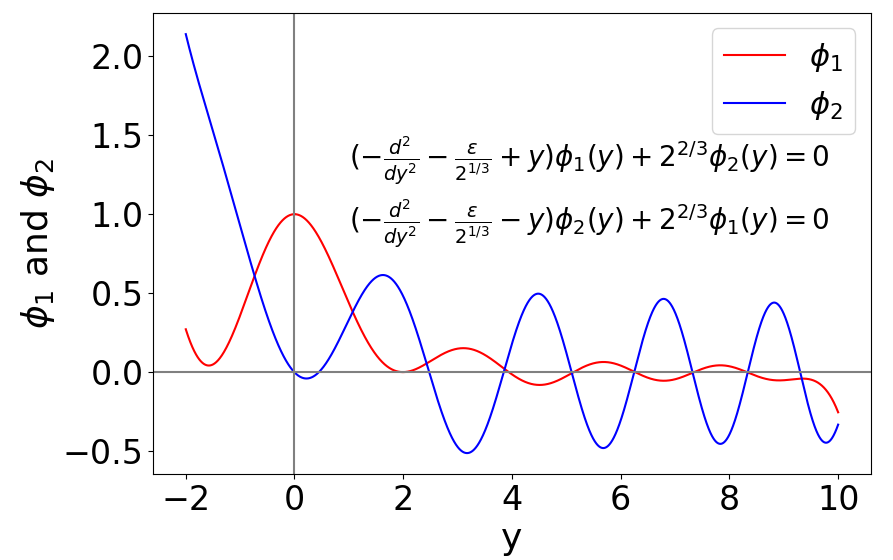}
    \includegraphics[height=3.3cm]{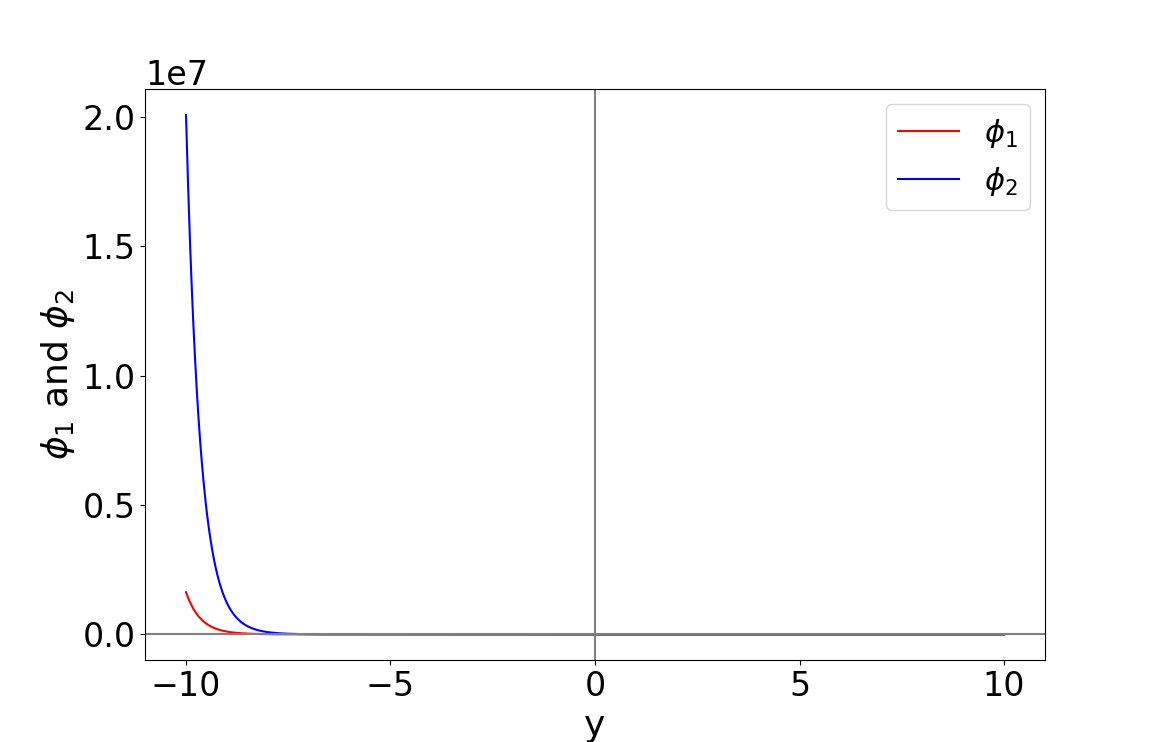}
\caption*{Fig. 9. Left: Original result (Ref. \cite{bialynicki-birulaTimeCrystalsMade2021}) with parameter $y$ ranging from $-10$ to $10$. Middle: Reproduced result with a detailed view of parameter $y$ ranging from $-2$ to $10$. Right: Reproduced result showing the full range of parameter $y$ from $-10$ to $10$, matching the original figure’s parameter range.}
\end{figure}

The eigenfunction components $\phi_1(y)$ and $\phi_2(y)$ correspond to the eigenvalue $\varepsilon=2$ of the quantum Hamiltonian (19), with initial values at $y=0$: $\phi_1=1$, $\phi_1'=0$, $\phi_2=0$, and $\phi_2'=-0.354651985$. 
It should be noted, however, that Figure 9 in Ref. \cite{bialynicki-birulaTimeCrystalsMade2021} clearly shows $\phi_2'(0)>0$ (highlighted in green boxes). Moreover, in the $y<0$ region--particularly at $y=-10$--the magnitude of $\phi_2$ reaches up to $2\times10^7$, which is in direct contradiction to the description provided in Ref. \cite{bialynicki-birulaTimeCrystalsMade2021}.

\begin{figure}[H]
    \centering
    \includegraphics[height=3cm]{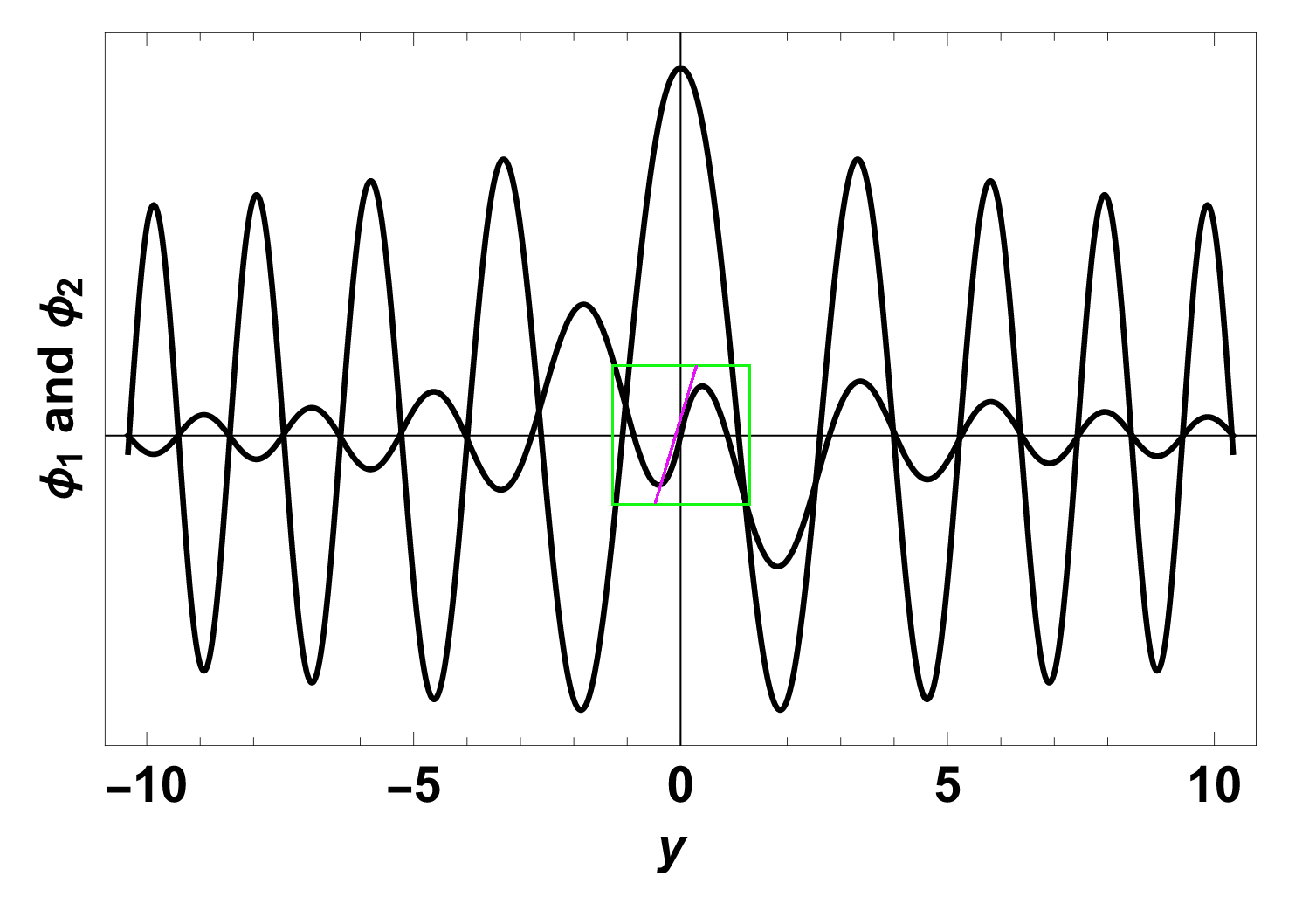}
    \includegraphics[height=3cm]{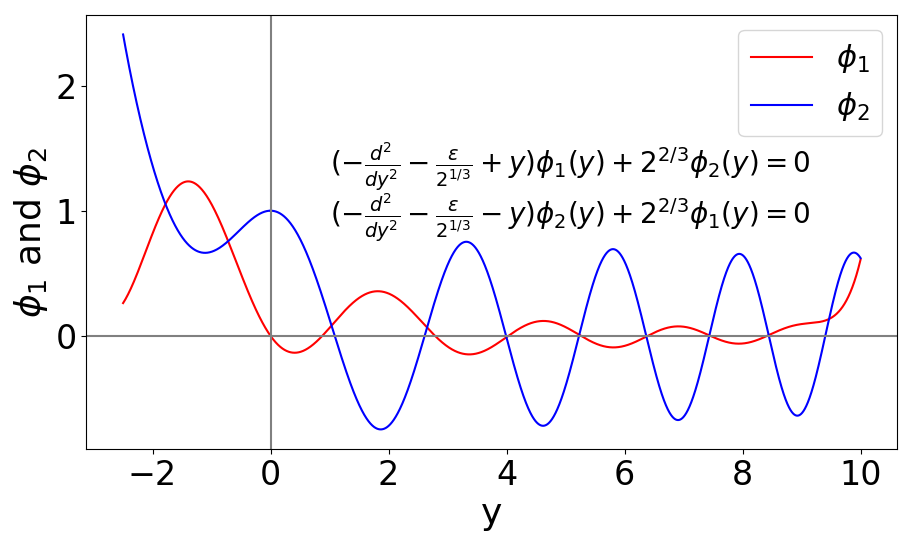}
    \includegraphics[height=3.3cm]{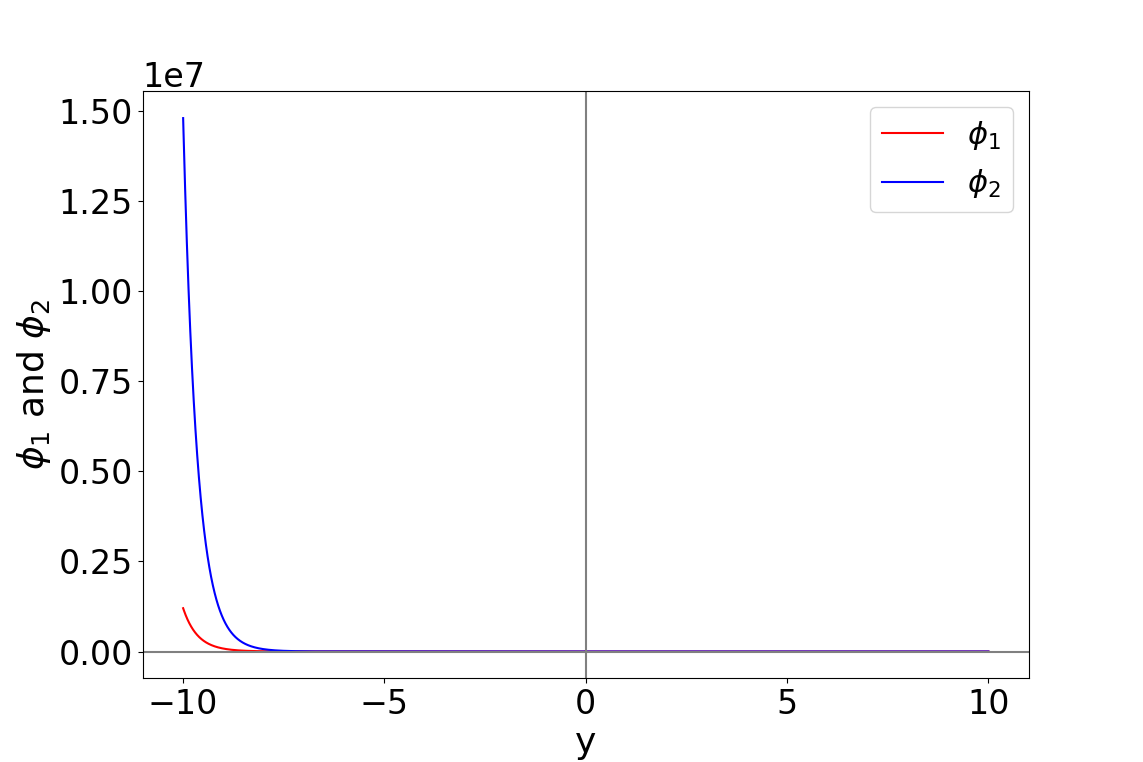}
\caption*{Fig. 10. Left: Original result (Ref. \cite{bialynicki-birulaTimeCrystalsMade2021}) with parameter $y$ ranging from $-10$ to $10$. Middle: Reproduced result with a detailed view of parameter $y$ ranging from $-2$ to $10$. Right: Reproduced result showing the full range of parameter $y$ from $-10$ to $10$, matching the original figure’s parameter range.}
\end{figure}

The eigenfunction components $\phi_1(y)$ and $\phi_2(y)$ correspond to the eigenvalue $\varepsilon=2$ of the quantum Hamiltonian (19), with initial values at $y=0$: $\phi_1=0$, $\phi_1'=-0.665192338$, $\phi_2=1$, and $\phi_2'=0$.
Figures 10 and 9 share the same issue, that is, Figure 10 in reference \cite{bialynicki-birulaTimeCrystalsMade2021} shows $\phi_1'(0)>0$(highlighted in green boxes) and attention is also drawn to the fact that, at $y=-10$, the magnitude of $\phi_2$ reaches $1.5\times10^7$, which remains inconsistent with the description provided in Ref. \cite{bialynicki-birulaTimeCrystalsMade2021}.

\begin{figure}[H]
    \centering
    \includegraphics[height=3cm]{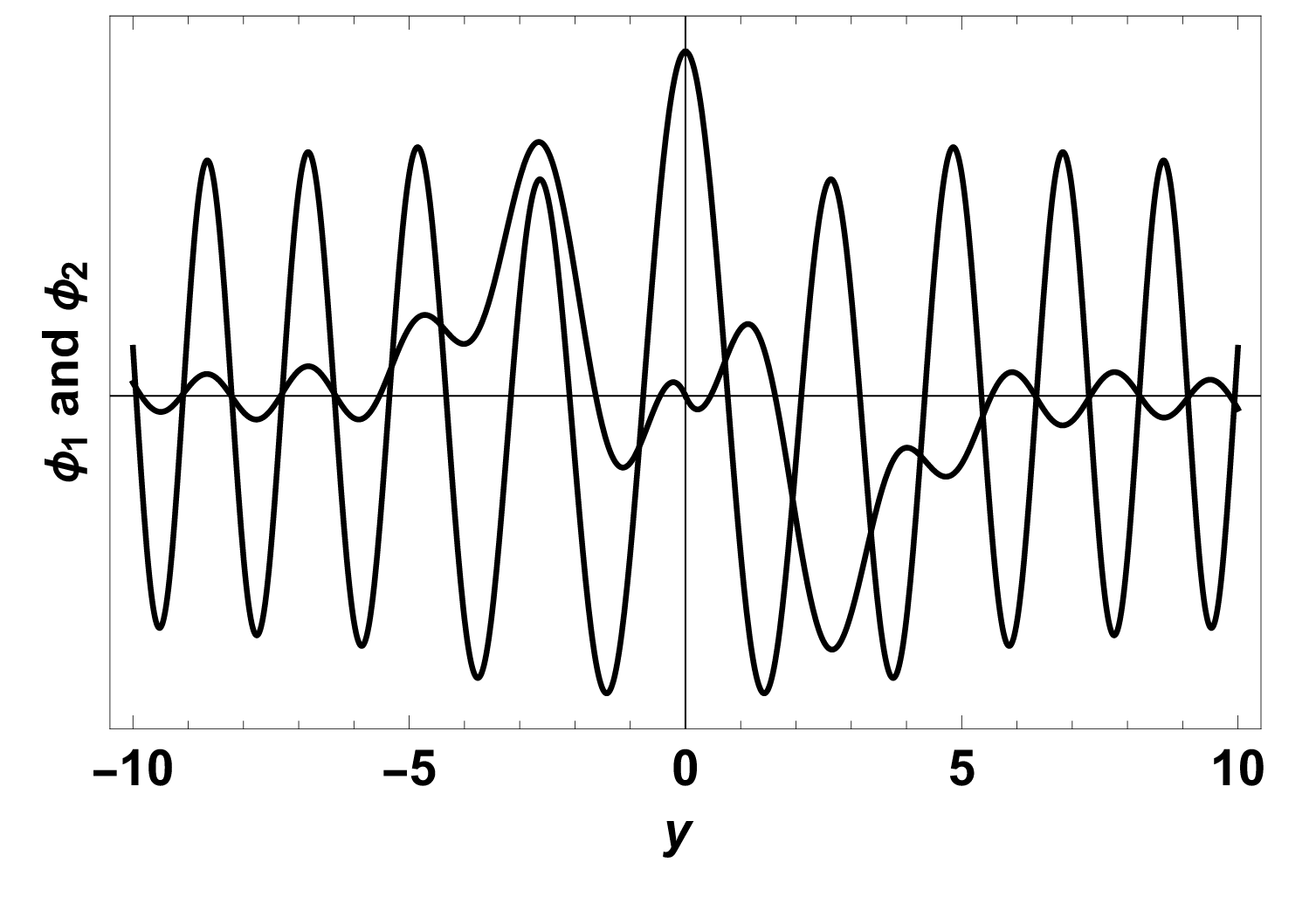}
    \includegraphics[height=3cm]{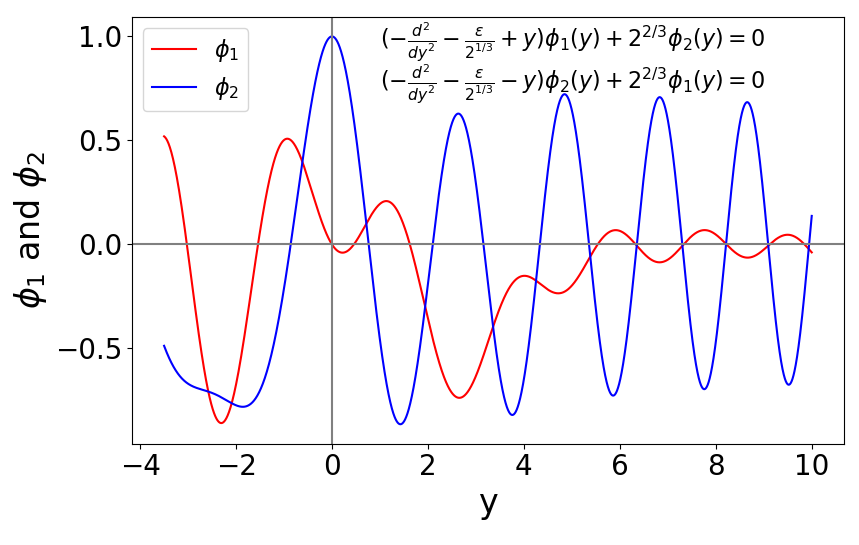}
    \includegraphics[height=3.3cm]{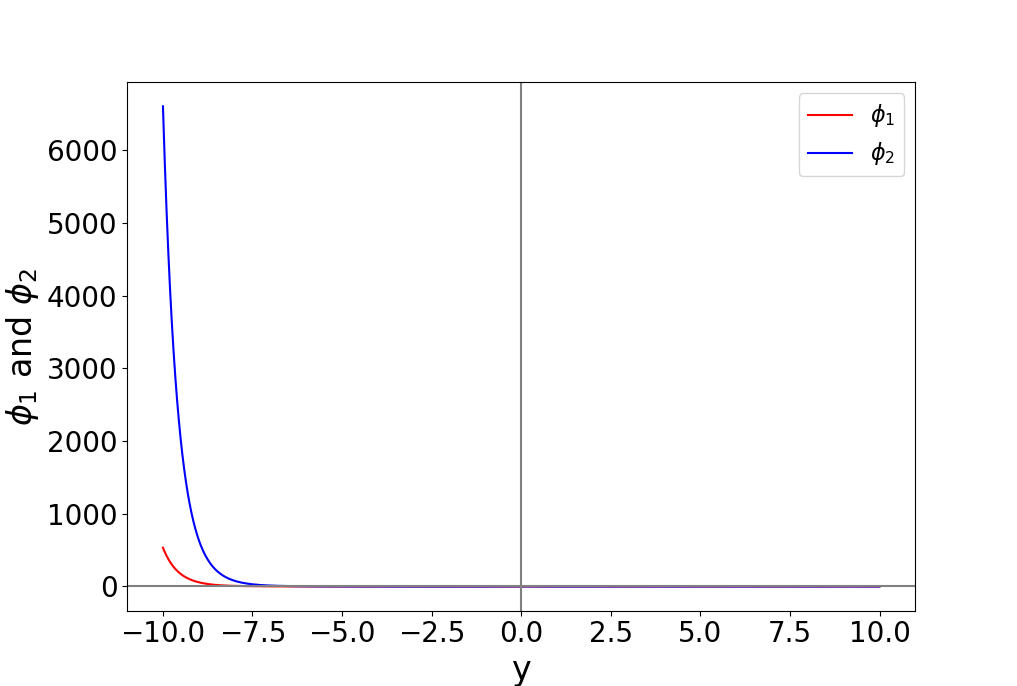}
\caption*{Fig. 11. Left: Original result (Ref. \cite{bialynicki-birulaTimeCrystalsMade2021}) with parameter $y$ ranging from $-10$ to $10$. Middle: Reproduced result with a detailed view of parameter $y$ ranging from $-4$ to $10$. Right: Reproduced result showing the full range of parameter $y$ from $-10$ to $10$, matching the original figure’s parameter range.}
\end{figure}

The eigenfunction components $\phi_1(y)$ and $\phi_2(y)$ correspond to the eigenvalue $\varepsilon=5$ of the quantum Hamiltonian (19), with initial values at $y=0$: $\phi_1=0$, $\phi_1'=-0.36012$, $\phi_2=1$, and $\phi_2'=0$.
Figure 11 shows that $\phi_2$ exceeds 6000 near $y=-10$, which is significantly different from the original figure in Ref. \cite{bialynicki-birulaTimeCrystalsMade2021}. However, we can obtain the figure from the original paper by transforming our results in the following ways:
\begin{enumerate}
    \item Ignore the numerical results for $y<0$.
    \item A 180-degree clockwise rotation of the red $\phi_1(y)$ from the right figure about the origin results in the $\phi_1(y)$ shown in Figure 11.
    \item Mirroring the blue $\phi_2(y)$ from the right figure across the line $y=0$ will produce the $\phi_2(y)$ in the original figure.
\end{enumerate}

To resolve the discrepancies between the reproduced results and those reported in the original work, we suggest further clarification of the numerical implementation, which would help enhance the consistency between theoretical predictions and graphical results.

\section{Theoretical Origins of Quantum Solution Inconsistencies}

\subsection{Parity Symmetry Violation: A Mathematical Necessity}

The original Eq. (21) from reference \cite{bialynicki-birulaTimeCrystalsMade2021} reads:

\begin{equation}
    \begin{cases}
        \left(-\frac{d^2}{dy^2} - \frac{\mathcal{E}}{2^{1/3}} + y\right)\phi_1(y) + 2^{2/3}\phi_2(y) = 0, \\
        \left(-\frac{d^2}{dy^2} - \frac{\mathcal{E}}{2^{1/3}} - y\right)\phi_2(y) + 2^{2/3}\phi_1(y) = 0.
    \end{cases}
    \label{eq:original}
\end{equation}

As clearly demonstrated in Figures 9--11 of Ref. \cite{bialynicki-birulaTimeCrystalsMade2021}, the eigenfunctions $\phi_1(y)$ and $\phi_2(2)$ distinctly exhibit odd and even parity, respectively.
Assuming $\phi_1(y)$ is odd and $\phi_2(y)$ is even, substituting $-y$ for $y$ in Eq. \eqref{eq:original} yields:

\begin{equation}
    \begin{cases}
        \left(-\frac{d^2}{dy^2} - \frac{\mathcal{E}}{2^{1/3}} - y\right)\phi_1(y) - 2^{2/3}\phi_2(y) = 0, \\
        \left(-\frac{d^2}{dy^2} - \frac{\mathcal{E}}{2^{1/3}} + y\right)\phi_2(y) - 2^{2/3}\phi_1(y) = 0.
    \end{cases}.
    \label{eq:original1}
\end{equation}

For equations governing parity-definite solutions, invariance under $y\to-y$ is expected \cite{arfken1985mathematical}. However, Eqs. \eqref{eq:original} and \eqref{eq:original1} differ in structure, suggesting that solutions $\phi_1(y)$ and $\phi_2(y)$ may not strictly adhere to the parity properties implied by the figures. 

Now, from the perspective of linear superposition, we further demonstrate that Equation (21) does not possess parity symmetry. By substituting $y$ with $-y$ in Eq. \eqref{eq:original}, we obtain:

\begin{equation}
    \begin{cases}
        \left(-\frac{d^2}{dy^2} - \frac{\mathcal{E}}{2^{1/3}} - y\right)\phi_1(-y) + 2^{2/3}\phi_2(-y) = 0, \\
        \left(-\frac{d^2}{dy^2} - \frac{\mathcal{E}}{2^{1/3}} + y\right)\phi_2(-y) + 2^{2/3}\phi_1(-y) = 0.
    \end{cases}
    \label{eq:original2}
\end{equation}
This implies that $\phi_2(-y)$ and $\phi_1(-y)$ are also solutions to Eq. \eqref{eq:original}. If $\phi_1(y)$ and $\phi_2(-y)$ are linearly dependent, without loss of generality, assume $\phi_1(y)=k\phi_2(-y)$, where $k$ is a constant. Then Eq. \eqref{eq:original2} reduces to
\begin{equation}
    \begin{cases}
        \left(-\frac{d^2}{dy^2} - \frac{\mathcal{E}}{2^{1/3}} - y\right)\phi_1(-y) + \frac{2^{2/3}}{k}\phi_1(y) = 0, \\
        \left(-\frac{d^2}{dy^2} - \frac{\mathcal{E}}{2^{1/3}} + y\right)\phi_1(y) + 2^{2/3}k\phi_1(-y) = 0.
    \end{cases}
    \label{eq:original2a}
\end{equation}
If $\phi_1(y)$ is an odd function, then by adding the two equations in Eq. \eqref{eq:original2a}, we obtain
\begin{equation}
    \left[ 2y+2^{2/3}(\frac{1}{k}-k) \right]\phi_1(y)=0.
    \label{eq:original2b}
\end{equation}
Due to the arbitrariness of the independent variable $y$, it must hold that
\begin{equation}
    \phi_1(y)=0.
    \label{eq:original2c}
\end{equation}
If $\phi_1(y)$ is an even function, then by subtracting the two equations in Eq \eqref{eq:original2a}, we still arrive at the conclusion given in Eq \eqref{eq:original2c}. If $\phi_1(y)$ and $\phi_2(-y)$ are linearly independent, then any solution of Eq. \eqref{eq:original} can be expressed as a linear combination \cite{arfken1985mathematical}, such as
\begin{equation}
    \varphi_1(y)=a\phi_1(y)+b\phi_2(-y),\quad
    \varphi_2(y)=a\phi_2(y)+b\phi_1(-y).
    \label{eq:original3}
\end{equation}
If $\varphi_1(y)$ is an odd function and $\varphi_2(y)$ is an even function, then
\begin{equation}
    \begin{cases}
        a\left[ \phi_1(-y)+\phi_1(y) \right]+b\left[ \phi_2(y)+\phi_2(-y) \right]=0,\\
        a\left[ \phi_2(-y)-\phi_2(y) \right]+b\left[ \phi_1(y)-\phi_1(-y) \right]=0.
    \end{cases}
    \label{eq:original4}
\end{equation}
Due to the arbitrariness of $\phi_1(y)$ and $\phi_2(y)$, it must be that $a=b=0$. If $\varphi_1(y)$ is an even function and $\varphi_2(y)$ is an odd function, then by the same reasoning, we obtain $a=b=0$. Therefore, it can be concluded that the solutions of Eq. \eqref{eq:original} do not possess even or odd symmetry.

The above mathematical analysis shows that the solutions $\phi_1$ and $\phi_2$ of Eq. (21) do not exhibit parity symmetry, which is consistent with our numerical reproduction results. However, Figures 9--11 in Ref. \cite{bialynicki-birulaTimeCrystalsMade2021} display clear parity symmetry in both $\phi_1$ and $\phi_2$.

\subsection{Unphysical Basis Transformation: Amplifying the Artifact}

The primary concern regarding the original study stems from the apparent lack of theoretical foundation for the transformation proposed in Equation (20). Specifically, it remains unclear why the solutions $\Psi_1(y)$ and $\Psi_2(y)$ derived from Equation (19) would necessarily permit decomposition into the particular form presented in Equation (20).

By substituting the Pauli matrices \cite{pauli1927zusammenhang} into Equation (19) of Ref. \cite{bialynicki-birulaTimeCrystalsMade2021}, we obtain
\begin{equation}
    \mqty[-\frac{1}{2}\dv[2]{x}-\varepsilon & 2(1-ix) \\2(1+ix) & -\frac{1}{2}\dv[2]{x}-\varepsilon  ]\mqty[\Psi_1 \\ \Psi_2]=0.
    \label{eq:sjjtx1}
\end{equation}
Following Ref. \cite{bialynicki-birulaTimeCrystalsMade2021}, applying the transformation $x=2^{-2/3}y$, we get
\begin{equation}
    \begin{cases}
        \left( -\dv[2]{y}-\frac{\varepsilon}{2^{1/3}} \right)\Psi_1(y)+\left( 2^{2/3}-iy \right)\Psi_2(y)=0,\\
        \left( -\dv[2]{y}-\frac{\varepsilon}{2^{1/3}} \right)\Psi_2(y)+\left( 2^{2/3}+iy \right)\Psi_1(y)=0.
    \end{cases}
    \label{eq:sjjtx2}
\end{equation}
Plugging Eq. (20) from Ref. \cite{bialynicki-birulaTimeCrystalsMade2021} into Eq. \eqref{eq:sjjtx2} yields
\begin{equation}
    \begin{cases}
        \left( -\dv[2]{y}-\frac{\varepsilon}{2^{1/3}}+y \right)\phi_1(y)+2^{2/3}\phi_2(y)+i\left[ \left( -\dv[2]{y}-\frac{\varepsilon}{2^{1/3}}-y \right)\phi_2(y)+2^{2/3}\phi_1(y) \right]=0,\\
        \left( -\dv[2]{y}-\frac{\varepsilon}{2^{1/3}}-y \right)\phi_2(y)+2^{2/3}\phi_1(y)+i\left[ \left( -\dv[2]{y}-\frac{\varepsilon}{2^{1/3}}+y \right)\phi_1(y)+2^{2/3}\phi_2(y) \right]=0.
    \end{cases}
    \label{eq:sjjtx3}
\end{equation}
By setting the real and imaginary parts of Eq. \eqref{eq:sjjtx3} to zero separately, we can obtain
\begin{equation}
    \begin{cases}
        \left( -\dv[2]{y}-\frac{\varepsilon}{2^{1/3}}+y \right)\phi_1(y)+2^{2/3}\phi_2(y)=0,\\
        \left( -\dv[2]{y}-\frac{\varepsilon}{2^{1/3}}-y \right)\phi_2(y)+2^{2/3}\phi_1(y)=0,
    \end{cases}
    \label{eq:sjjtx3a}
\end{equation}
which reduces to Eq. (21) in reference \cite{bialynicki-birulaTimeCrystalsMade2021}. Based on the detailed calculations above, it can be further confirmed that the functions $\phi_1(y)$ and $\phi_2(y)$ appearing in Eq. (20) of Ref. \cite{bialynicki-birulaTimeCrystalsMade2021} are real-valued.
Based on Eq. (20) in reference \cite{bialynicki-birulaTimeCrystalsMade2021}, we have
\begin{equation}
    \psi_1(y) = \frac{1}{\sqrt{2}}[\phi_1(y) + i\phi_2(y)], \quad \psi_2(y) = \frac{1}{\sqrt{2}}[\phi_2(y) + i\phi_1(y)],
    \label{eq:basis20}
\end{equation}
It follows that
\begin{equation}
    i\Psi^{*}_2(y)=\Psi_1(y).
    \label{eq:basis20a}
\end{equation}
Simultaneously, by taking the complex conjugate of the second expression in Eq. \eqref{eq:sjjtx2}, the following result is obtained:
\begin{equation}
    \begin{cases}
        \left( -\dv[2]{y}-\frac{\varepsilon}{2^{1/3}} \right)\Psi_1(y)+\left( 2^{2/3}-iy \right)\Psi_2(y)=0\\
        \left( -\dv[2]{y}-\frac{\varepsilon}{2^{1/3}} \right)\Psi^{*}_2(y)+\left( 2^{2/3}-iy \right)\Psi^{*}_1(y)=0
    \end{cases}.
    \label{eq:sjjtx4}
\end{equation}

While $i\Psi^{*}_2(y)=\Psi_1(y)$ ensures that Eq. \eqref{eq:sjjtx4} is satisfied, it is not the only condition that can satisfy Eq. \eqref{eq:sjjtx4}. Therefore, $i\Psi^{*}_2(y)=\Psi_1(y)$ is a sufficient but not necessary condition for Eq. \eqref{eq:sjjtx4} to hold. However, this is merely a mathematical conclusion. Since Eq. \eqref{eq:sjjtx4} is a physical equation, we have not yet taken sufficient physical considerations into account to justify that its solutions must be decomposed in the form $i\Psi^{*}_2(y)=\Psi_1(y)$. This is one of my central concerns.

%
%

\section{Acknowledgments}

The author expresses deep respect for the pioneering contributions of Prof. Bialynicki-Birula to quantum theory and acknowledges the intellectual foundation provided by Ref. \cite{bialynicki-birulaTimeCrystalsMade2021}.

\section*{Competing Interests}

The author declares no competing interests.

\bibliographystyle{unsrt}
\bibliography{reference}

\begin{thebibliography}{1}
\providecommand{\url}[1]{\texttt{#1}}
\providecommand{\urlprefix}{URL }
\expandafter\ifx\csname urlstyle\endcsname\relax
  \providecommand{\doi}[1]{doi:\discretionary{}{}{}#1}\else
  \providecommand{\doi}{doi:\discretionary{}{}{}\begingroup
  \urlstyle{rm}\Url}\fi
\providecommand{\eprint}[2][]{\url{#2}}

\bibitem{bialynicki-birulaTimeCrystalsMade2021}
I.~Bialynicki-Birula and Z.~Bialynicka-Birula,
\newblock \emph{Time crystals made of electron-positron pairs},
\newblock Phys. Rev. A \textbf{\textbf{104}}, 022203 (2021),
\newblock \doi{10.1103/PhysRevA.104.022203}.

\bibitem{arfken1985mathematical}
G.~B. Arfken,
\newblock \emph{Mathematical Methods for Physicists},
\newblock Academic Press, San Diego, USA, 3 edn. (1985).

\bibitem{pauli1927zusammenhang}
W.~Pauli,
\newblock \emph{{\"U}ber den zusammenhang des abschlusses der elektronengruppen
  im atom mit der komplexstruktur der spektren},
\newblock Zeitschrift f{\"u}r Physik \textbf{43}(9-10), 601 (1927),
\newblock \doi{10.1007/BF01900315}.

\end{thebibliography}

\end{document}